\documentclass[aps,amsmath,amssymb,nofootinbib,reprint]{revtex4-1}

\usepackage{graphicx}
\usepackage{dcolumn}
\usepackage{bm}
 \usepackage{url}
 \usepackage[utf8]{inputenc}
\usepackage[T1]{fontenc}
\usepackage{mathptmx}
\usepackage{gensymb}
\usepackage{etoolbox}
 \usepackage{color}
 \usepackage{lipsum}
 \usepackage[normalem]{ulem} 
 
\begin{document}

\preprint{PRD}

\title{Sensitivity of ultralight axion dark matter search with optical quantum sensors}
\author{Young Jin Kim$^{\mathsection}$}
\email{youngjin@lanl.gov.}
\author{Leanne Duffy$^{\mathsection}$}
\email{ldd@lanl.gov.}
\author{Igor Savukov$^{\mathsection}$}
\email{isavukov@lanl.gov.}
\author{Ping-Han Chu}

\affiliation{Los Alamos National Laboratory, P.O. Box 1663, Los Alamos, New Mexico 87545, USA}
\def\thefootnote{$\mathsection$}\footnotetext{these authors are co-first authors who contributed equally to this work.}
\date{\today}

\begin{abstract}

An optical quantum sensor (OQS) based on lasers and alkali-metal atoms is a sensitive ambient-temperature magnetometer that can be used in axion dark matter search with an inductor-capacitor (LC) circuit at kHz and MHz frequencies. We have previously investigated the sensitivity of an LC circuit-OQS axion detector to ultralight axion dark matter that could be achieved using a fT-noise OQS constructed in our lab. In this paper, we investigate the sensitivity that could be potentially reached by an OQS performing close to the fundamental quantum noise levels of 10~aT/$\sqrt{\text{Hz}}$. To take advantage of the quantum-limited OQS, the LC circuit has to be made of a superconductor and cooled to low temperature of a few K.
After considering the intrinsic noise of the advanced axion detector and characterizing possible background noises, we estimate that such an experiment could probe  benchmark QCD axion models in an unexplored mass range near 10~neV. Reaching such a high sensitivity is a difficult task, so we have conducted some preliminary experiments with a large-bore magnet and a prototype axion detector consisting of a room-temperature LC circuit and a commercial OQS unit.
This paper describes the prototype experiment and its projected sensitivity to axions in detail.
\end{abstract}

\maketitle

\section{Introduction}
Several mysteries in particle and astrophysics suggest that there are new particles yet to be discovered. One of them is an elusive cosmic substance, six times more abundant than the ordinary
matter in the Universe, known as dark matter~\cite{doi:10.1073/pnas.1516944112}. Another, seemingly unrelated mystery, is the fact that the strong nuclear interactions, described by quantum chromodynamics (QCD), are invariant under time-reversal with $10^{-10}$ precision or better. The QCD axion, a hypothetical particle first proposed in the 1970s~\cite{PhysRevLett.38.1440,PhysRevD.16.1791,PhysRevLett.40.279,PhysRevLett.40.223}, is an excellent candidate for the Universe’s dark matter~\cite{Duffy_2009}: if lighter than $\sim$meV in mass, it can be produced with the correct abundance and temperature in the early Universe to account for dark matter. Furthermore, it provides a dynamical mechanism to suppress time-reversal asymmetries in QCD~\cite{PhysRevLett.38.1440,PhysRevD.16.1791,PhysRevLett.40.279,PhysRevLett.40.223}. 

The target of many experimental designs for axion direct detection is the extremely small signal induced by the axion's weak coupling to electromagnetism~\cite{PhysRevLett.120.151301,PhysRevLett.112.131301,PhysRevLett.127.081801,SHAFT}. To date, the benchmark QCD axion models of Dine-Fischler-Srednicki-Zhitnitsky (DFSZ)~\cite{DINE1981199,osti_7063072} and Kim-Shifman-Vainshtein-Sakharov (KSVZ)~\cite{PhysRevLett.43.103,SHIFMAN1980493} have both been probed in masses above 2.66~$\mu$eV by the Axion Dark Matter eXperiment (ADMX)~\cite{PhysRevLett.120.151301} using a resonant cavity haloscope~\cite{PhysRevLett.51.1415,PhysRevD.36.974}. The haloscope relies on the interaction of the dark matter axion field with a strong static magnetic field generated from a superconducting magnet. The smallest axion mass of $\sim\mu$eV  that ADMX can search is limited by the physical size of the resonant cavity ($\sim1$~m), and thus the magnet bore in which it must fit: the cavity size should be comparable to the axion Compton wavelength, for example, $\sim1$~m and $\sim1$~km for axion mass of $\mu$eV and neV, respectively. 

Axions with mass around 10~neV are predicted by Grand Unified Theory (GUT) models of particle physics~\cite{Langacker:2012}.  Due to the physical lower mass bound, cavity haloscopes cannot be used to search for these ultralight axions.  A recent proposal to extend the use of cavity searches in ADMX is the use of reentrant cavities to reach lower masses, down to around 0.4~$\mu$eV~\cite{chakrabarty2023low}.  At axion masses below this, a different experimental approach is required. As a part of that, an axion detector using an inductor-capacitor (LC) circuit coupled to a superconducting quantum interference device (SQUID), as a senstive magnetometer, was first proposed by Sikivie, Sullivan, and Tanner~\cite{PhysRevLett.112.131301}.  We have previously investigated using an optical quantum sensor (OQS) as a sensitive magnetometer~\cite{PhysRevD.97.072011} and have developed an experiment prototype based on a commercial OQS unit. During this prototype development, we realized that the detector sensitivity can be potentially improved with a quantum-limited OQS, with a field noise floor of the order of 10~aT/$\sqrt{\text{Hz}}$. In this paper, we study the sensitivity of an advanced LC circuit-OQS experiment designed to take full advantage of the noise of a quantum-limited OQS.

This study is organized as follows.  In Section~\ref{sec:signal}, we discuss the magnetic signature of dark matter axions in an LC circuit-OQS axion detector. 
In Section~\ref{sec:noise}, we discuss noise sources in  axion detection experiments. We investigate the intrinsic noise of the axion detector arising from the intrinsic noise of the OQS and the thermal Johnson noise of the LC circuit. For the improved axion detector sensitivity with a quantum-limited OQS, the LC circuit must be cooled to low temperature of a few K; otherwise the thermal noise of the circuit at room temperature dominates. In this case, other background noises could become significant. Therefore, we also investigate possible background noises, including the backaction noise of the OQS on the LC circuit and the thermal noise of surrounding magnetic shield.  In Section~\ref{sec:exp}, we describe development of a prototype experiment with a room temperature LC circuit and a commercial fT-noise OQS. The sensitivity of this prototype setup to axions is discussed in Section~\ref{sec:sens}. 
Potential improvements to take advantage of a quantum-limited OQS and the advanced axion detector's sensitivity that can be achieved are discussed in Section~\ref{sec:imp}. We conclude our discussion in Section ~\ref{sec:conc}.

\section{Axion Signal and Detection Scheme}\label{sec:signal}

Dark matter axions behave like a classically field oscillating at the axion Compton frequency, $\omega_a=m_a$ with $m_a$ being the axion mass, and permeating the entire Universe~\cite{https://doi.org/10.48550/arxiv.1707.04591}. Thus the axion field can be written as $a(t)=a_0\sin{(m_a t)}$, where $a_0$ is the local amplitude of the axion field. Here natural units  $c=\hbar=\mu_0=1$ are used. The existence of axions in the presence of a static magnetic field $\vec{B_0}$ gives rise to additional terms in the Maxwell equations~\cite{PhysRevLett.112.131301}. In particular, Ampere's Law is modified to
\begin{align}
    \vec{\nabla}\times\vec{B_0}-\frac{\partial\vec{E}}{\partial{t}}=& g\Big(\vec{E}\times\vec{\nabla}a-\vec{B_0}\frac{\partial a}{\partial t}\Big)+\vec{j}_{e},
    \label{eq:modifiedmaxwell}
\end{align}
where $g$ is the coupling constant of the axion to two photons~\cite{SREDNICKI1985689} and $\vec{j}_{e}$ is the electrical current density associated with ordinary matter. It follows that the homogeneous axion field ($\vec{\nabla}a\approx0$) can induce an electrical current density along $\vec{B}_0$,
\begin{align}
\vec{j}_a(t) = -g\vec{B}_0[da(t)/dt]=-g\vec{B}_0\sqrt{2\rho_{DM}}\cos{(m_a t)},
\label{eq:Ja}
\end{align}
with the relation of $a_0=\sqrt{2\rho_{DM}}/m_a$, where $\rho_{DM}\approx0.3~ \text{GeV/cm}^3$ is the standard local dark matter density~\cite{patrignani:2017}.
In turn, $\vec{j}_a(t)$ can produce a minute perpendicular magnetic field $\vec{B}_a(t)$ oscillating at an angular frequency $m_a$ through $\vec{\nabla}\times \vec{B}_a = \vec{j}_a$. This led Sikivie, Sullivan, and Tanner to propose axion detection based on the LC circuit and SQUIDs ~\cite{PhysRevD.97.072011}.  Here we propose an LC circuit design with a gradiometer coil and a quantum-limited OQS for detection of the axion-induced oscillating magnetic field $\vec{B}_a$. 

\begin{figure}[t!]
\centering
\includegraphics[width=3.4in]{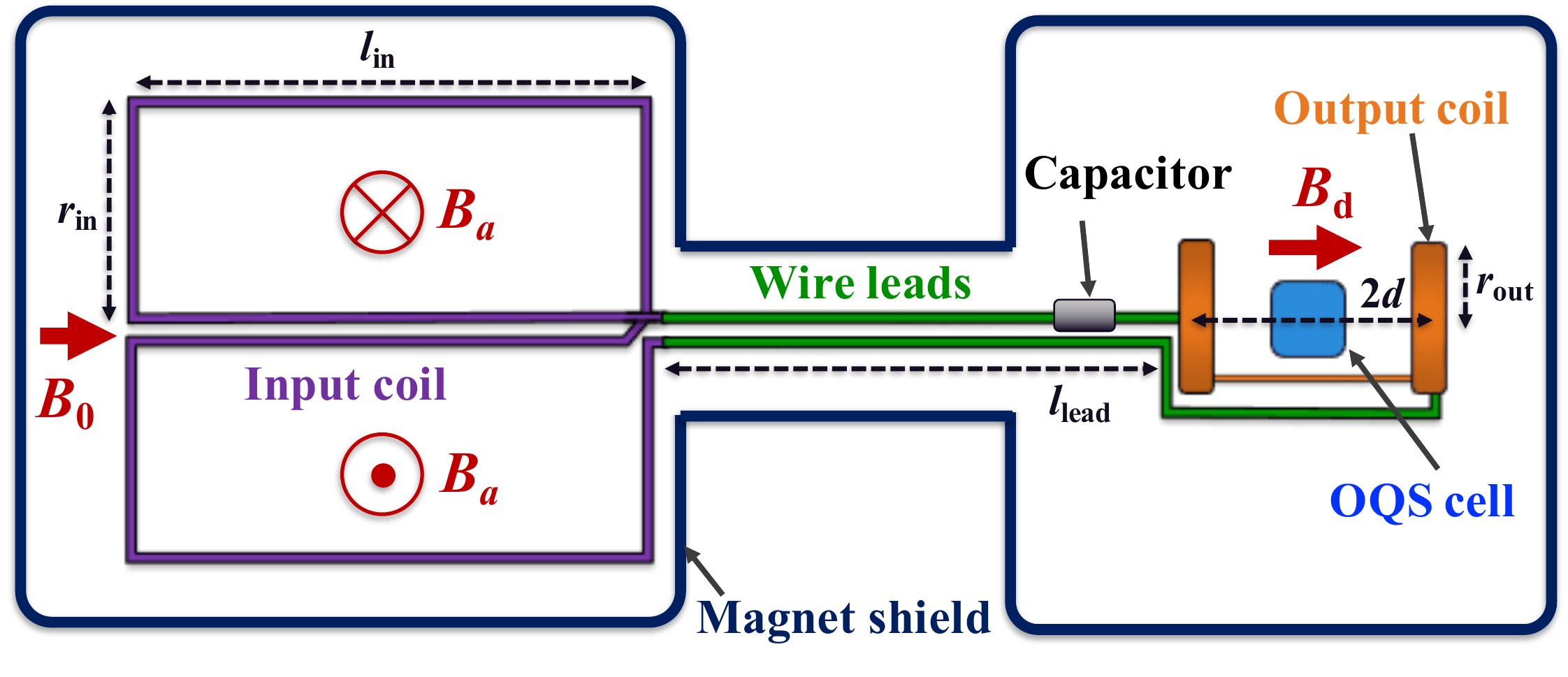}
\caption{Sketch of the LC circuit-OQS axion detector showing electrical connections of the LC circuit and geometrical arrangement of the OQS and magnetic shield (not to scale).}
\label{fig:config}
\end{figure}

A sketch of the axion detection concept with an OQS is shown in Fig.~\ref{fig:config}. The axion detector is comprised of two components. The first component is the first-order planar gradiometer input coil and a two-loop circular output coil, the two coils connected in series with capacitors (LC circuit) to resonantly amplify the $\vec{B}_a$. The second component is an OQS to detect the amplified magnetic field $\vec{B}_d$. 

\begin{figure}[t!]
\centering
\includegraphics[width=3.5in]{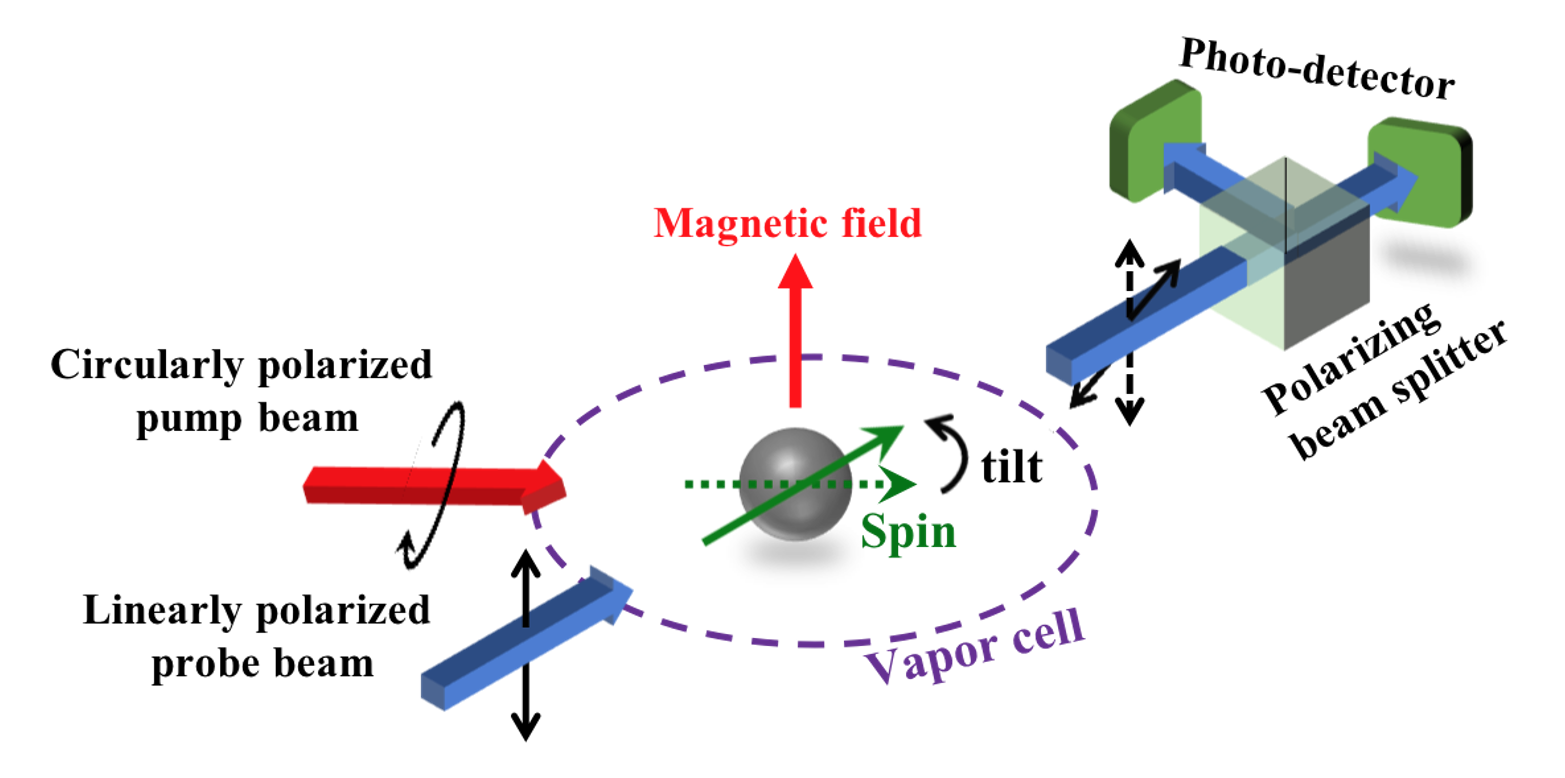}
\caption{OQS configuration: two laser beams overlap in a vapor cell of alkali-metal atoms. The circularly polarized pump beam orients atomic spins along its direction (the dashed green arrow). The external magnetic field tilts the spins by a small angle (the solid green arrow). The tilt leads to a rotation of the plane of linear polarization of the probe beam (dotted and solid black arrows), which is measured by a polarizing beam splitter and two photo-detectors.}
\label{fig:OQS}
\end{figure}

The OQS manipulates atomic spins for sensitive magnetic sensing based on lasers, alkali-metal vapor cells, and optical components~\cite{OPM-Budker}, and operates at ambient temperatures without the need for cryogens. Cryogen-free operation of the OQS has many advantages for various applications, and for the axion search the main advantage is convenience for OQS replacement and optimization. The basic principle for a typical implementation of OQSs is shown in Fig.~\ref{fig:OQS}. Typically, two laser beams are used: one circularly polarized pump beam to optically polarize the spins of unpaired electrons of alkali-metal atoms, such as rubidium (Rb) or potassium (K), and one linearly polarized probe beam to read out the state of the electron spins. The wavelengths of the laser beams are tuned to or near an atomic transition between the ground state and an excited state, most often the $P_{1/2}$ state. The laser beams are sent to overlap in an alkali-metal vapor cell heated to elevate an alkali-metal atom density, e.g., $\sim10^{14}$~cm$^{-3}$ in case of K atoms at a $\sim180~\degree$C cell temperature. The action of the pump beam, referred to as optical pumping, orients nearly all of the electron spins along its propagation direction. The interaction of a weak external magnetic field to be detected with the polarized electron spins leads to a change in the orientations of the spins. The degree of change is proportional to the strength of the magnetic field. The non-zero spin projection along the probe beam results in a rotation of the light linear polarization plane of the probe beam caused by the Faraday effect. This optical rotation is precisely detected with a polarizing beam splitter and two photo-detectors as a small difference in the balanced output. The OQS frequency of the maximum sensitivity, $\nu_m$, is tuned by a small bias static magnetic field $B_b$ through $\nu_m=\gamma B_b$, where $\gamma=7$ GHz/T is the gyromagnetic coefficient of Rb-87 or K-39/41 electron spins. 

To detect the amplified magnetic axion signal $\vec{B}_d$, the OQS vapor cell is placed at the center of the output coil. The input gradiometer coil is located inside a magnet bore at room temperature, producing a static magnetic field $B_0$. The $\vec{B}_a$ induces a voltage in the input gradiometer by Faraday’s law, which drives a current through the output coil producing an oscillating magnetic field $\vec{B}_d$ that is our observable axion signal. As $\vec{B}_a$ is azimuthally symmetric, each pickup loop of the input gradiometer covers each half-side of a horizontal plane through the magnet and central axis. This configuration doubles the $B_a$ signal, but greatly reduces the background magnetic noise that generates voltages of opposite signs. 

As this is a resonant LC circuit axion detector, many aspects are similar to the proposal of Sikivie, Sullivan, and Tanner~\cite{PhysRevLett.112.131301}, but an OQS is uniquely used as a sensitive magnetometer. Thus our design is modified for the coupling to the OQS.  Our initial experimental proposal is described in detail in a previous publication~\cite{PhysRevD.97.072011}. 

When the LC circuit resonates at an angular frequency equal to the axion mass, i.e., $\omega=1/\sqrt{(L_{\text{in}}+L_{\text{out}})C}=m_a$ with $L_{\text{in}}$ and $L_{\text{out}}$ being the inductances of the input and output coils and $C$ being the capacitance of the capacitor (this includes the two coils' self-capacitance and the capacitance of leads, as well as other parasitic effects), the magnitude of the current in the circuit is given by 
\begin{align}
|I| =\frac{Q|\Phi_a|}{L_{\text{in}}+L_{\text{out}}},
\label{eq:Axioncurrent}
\end{align}
where $Q=\omega(L_{\text{in}}+L_{\text{out}})/R$ is the quality factor of the LC circuit, $R$ is the total AC resistance of the circuit, and $\Phi_a$ is the axion-induced magnetic flux through the input gradiometer coil. Using Eq.~(\ref{eq:Ja}), cylindrical coordinates, $(z,\rho,\phi)$, and $\vec{B_{0}}=B_0\hat{z}$, the
axion-induced oscillating magnetic field $\vec{B}_a$ in Fig.~\ref{fig:config} is~\cite{PhysRevD.97.072011}
\begin{align}
\vec{B_{a}} =-\frac{g\sqrt{2\rho_{DM}}B_0\rho}{2}\hat{\phi},
\label{eq:ba_in}
\end{align}
and thus the magnitude of $\Phi_a$ is
\begin{align}
|\Phi_{a}|=|\int 2N_{\text{in}}\vec{B_{a}}\cdot {d\vec{A}}|=2 N_{\text{in}}\text{V}_{\text{in}}g\sqrt{2\rho_{DM}}B_0,
    \label{eq:phi_in}
\end{align}
where $N_{\text{in}}$ is the number of turns of each pickup loop of the input gradiometer coil and $\text{V}_{\text{in}}=l_{\text{in}}r_{\text{in}}^2 /4$ is a geometric factor for the input gradiometer coil with $l_{\text{in}}$ and $r_{\text{in}}$ as its length and width. The output coil is composed of two loops in series with the center-to-center spacing, $2d$. The current $I$ in Eq.~(\ref{eq:Axioncurrent}) flowing through the output coil produces the magnetic field $\vec{B}_d$ at the location of the OQS sensing volume and thus the magnitude of $\vec{B}_d$ is
\begin{align}
|B_{d}| =&\frac{N_{\text{out}}|I|}{r_{\text{out}}[1+(d/r_{\text{out}})^2]^{3/2}}\notag\\=&\frac{N_{\text{out}}Q|\Phi_a|}{r_{\text{out}}[1+(d/r_{\text{out}})^2]^{3/2}(L_{\text{in}}+L_{\text{out}})},
 \label{eq:bd_out1}
\end{align}
where $r_{\text{out}}$ and $N_{\text{out}}$ is the radius and the number of turns of each loop of the output coil, respectively. Substituting Eq.~(\ref{eq:phi_in}) to Eq.~(\ref{eq:bd_out1}), the field magnitude is
\begin{align}
    |B_d| = \frac{2N_{\text{in}}N_{\text{out}}Q\text{V}_{\text{in}} g \sqrt{2\rho_{DM}} B_0 }{r_{\text{out}}[1+(d/r_{\text{out}})^2]^{3/2}(L_{\text{in}}+L_{\text{out}})}.
    \label{eq:bd_out2}
\end{align}
This indicates that the large magnet bore size and strong field $B_0$ are critical to improve the axion signal strength.

As the axion mass (frequency) is unknown, the axion detector must be tuned across a range of frequencies. This will be achieved by simultaneous  tuning the OQS frequency range by changing its bias magnetic field and the LC circuit by adjusting capacitors.  When the LC circuit resonates at the axion mass, the $\vec{B}_d$ is enhanced by the quality factor of the LC circuit.  To find the axion, the task is to detect an extremely small signal above noise sources.  In the following, we investigate the noise contributions to determine the sensitivity that a quantum-limited OQS offers for axion dark matter detection.

\section{Noise Sources}\label{sec:noise}

\begin{figure}[t!]
\centering
\includegraphics[width=3.1in]{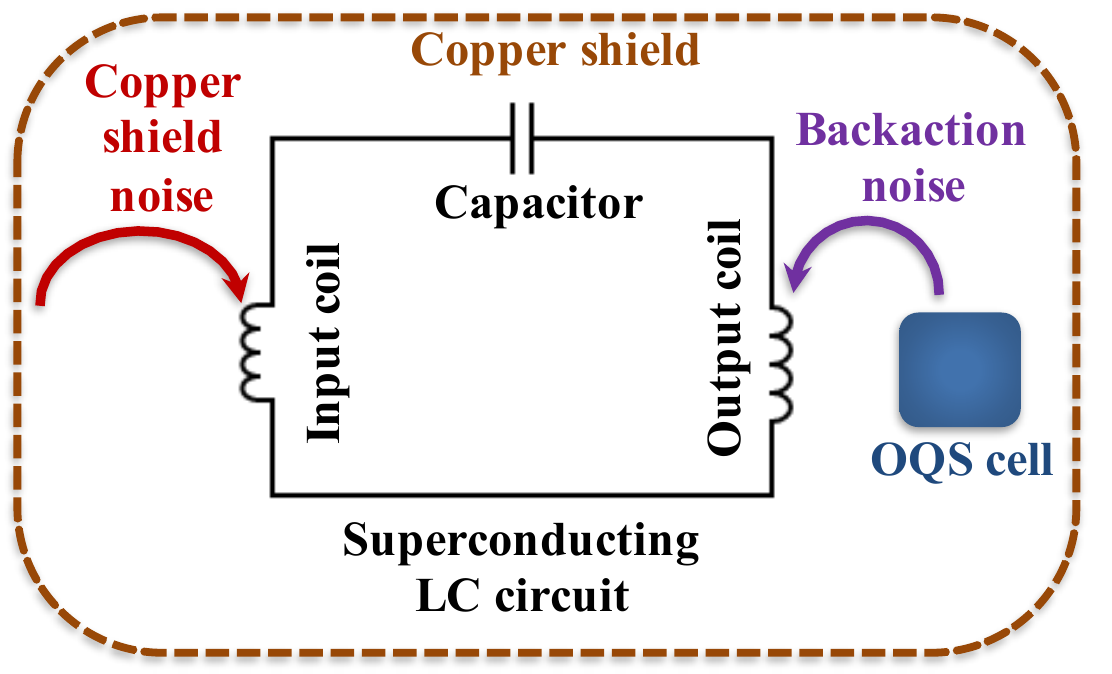}
\caption{Illustration of background noise contributions using an OQS as the magnetometer for axion detection with an LC circuit.  In addition to the intrinsic thermal (magnetic Johnson) noise of the LC circuit and the intrinsic field noise of the OQS, additional noise contributions come from the backaction noise of the OQS on the LC circuit and the thermal noise of experimental magnetic shielding, here copper shield.}
\label{fig:noise}
\end{figure}

The two main sources of noise in the axion detector in Fig.~\ref{fig:config} are the intrinsic magnetic field noise of the OQS, $\delta B_{\text{OQS}}$, and the magnetic Johnson noise (or thermal noise) of the LC circuit, $\delta B_{J}$.  Therefore, the total magnetic noise of this detector is given by
\begin{align}
\delta B_{d}=\sqrt{\delta B_{J}^2 +\delta B_{\text{OQS}}^2}.
   \label{eq:delBd}
\end{align}
When the field noise of the OQS approaches the fundamental quantum limit, the LC circuit must be cooled or its intrinsic thermal noise will dominate the experiment.  At low temperatures of the LC circuit, additional background noise sources must also be considered, which include the backaction noise of the OQS on the LC circuit and thermal noise from experimental magnetic shielding.  This is illustrated in Fig.~\ref{fig:noise}, and discussed in detail in the following.

\subsection{Fundamental Quantum Noise Limit of OQS}\label{subsec:QNL}

\begin{figure}[t!]
\centering
\includegraphics[width=3.5in]{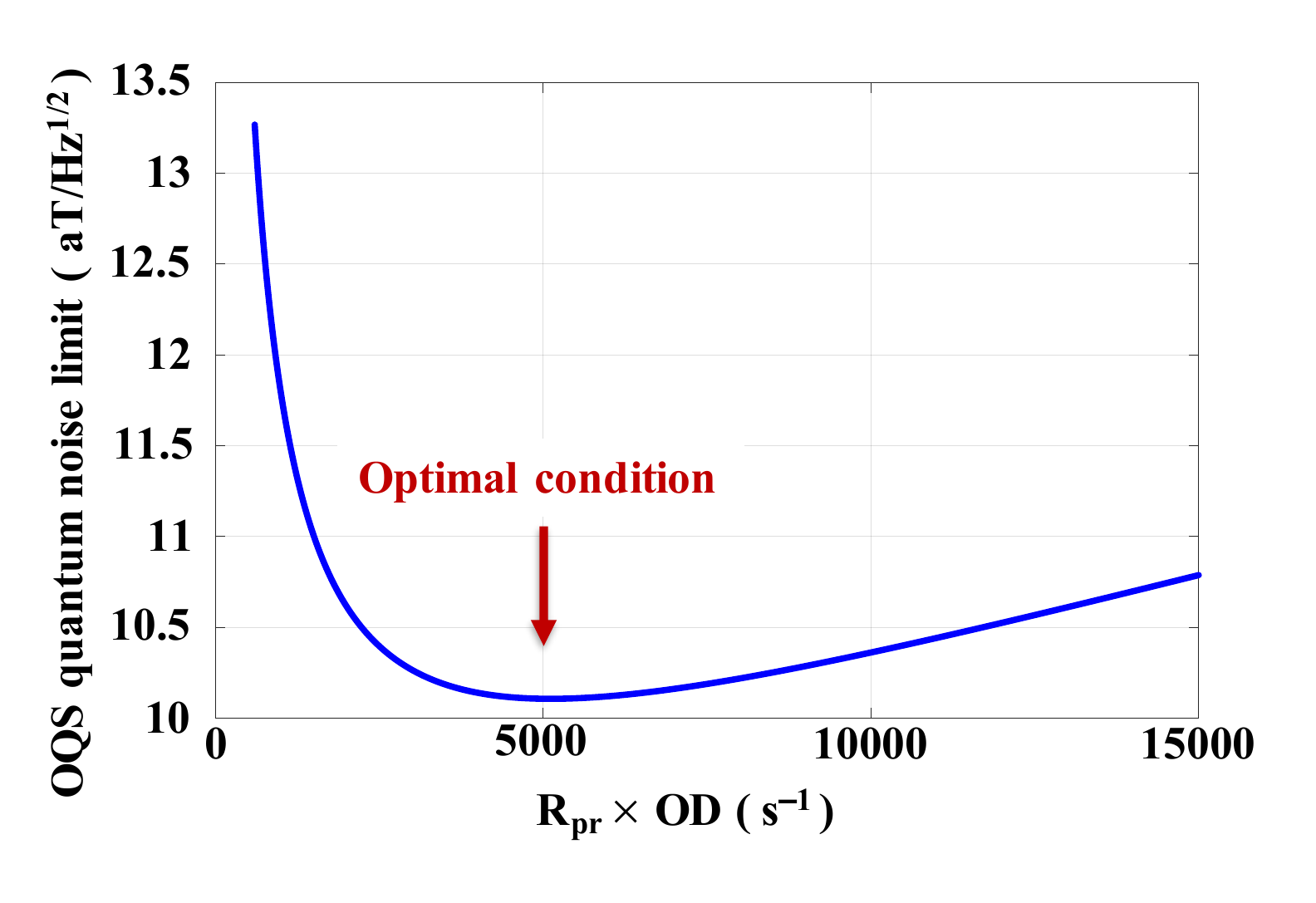}
\caption{Calculated fundamental quantum noise limit of OQS as a function of the product of $R_{pr}$ and OD.}
\label{fig:QNL}
\end{figure}

As in any quantum sensor, the intrinsic magnetic field noise of OQSs is ultimately limited by quantum fluctuations.  In this sensor, these are due to the finite number of alkali-metal atomic spins and the probe beam photons, both used for magnetic field sensing. Fundamentally, the three dominant sources of quantum field noise in OQSs are: (i) the spin projection noise caused by the finite number of alkali-metal atomic spins, (ii) the photon shot noise resulting from the finite number of probe beam photons, and (iii) the light-shift noise caused by fluctuations in the polarization of the probe beam~\cite{PhysRevLett.95.063004}. The quadrature sum of the three individual noise sources determines the fundamental quantum noise limit (QNL) of so-called radio-frequency OQS~\cite{PhysRevLett.95.063004}, 
\begin{align}
\delta B_{\text{QNL}}=\frac{1}{\gamma\sqrt{nV}}\sqrt{\frac{4}{T_2}+\frac{R_{pr}\text{OD}}{32}+\frac{8}{R_{pr}\text{OD}T_2^2\eta}}
\label{eq:QNL}
\end{align}
operating at frequencies much above 10~kHz where spin-exchange interaction is suppressed by optical pumping into the stretched state $|FF>$. Here $\gamma$  and $n$ are the gyromagnetic ratio and the density of alkali-metal atoms, respectively; $V$ is the active measurement cell volume defined by the overlap of the pump and probe beams; $T_2$ is the coherence time of the electron spins of alkali-metal atoms; $R_{pr}$ is the absorption rate of photons from the probe beam; $\eta\approx0.8$ is the photodiode quantum efficiency in the probe beam readout~\cite{PhysRevLett.95.063004}; and OD is the optical depth of the probe beam. If this expression inside the square root is optimized with respect to the product of $R_{pr}$ and OD (the optimal condition in Fig.~\ref{fig:QNL}), it will depend on $T_2$ as the first term, and the field noise of OQS will be limited by $T_2$ and the number of spins. Increasing $T_2$ can be obtained by pumping the electron spins into the stretched state to reduce the dominant spin-exchange relaxation by alkali-metal atom collisions~\cite{PhysRevLett.95.063004}. The number of spins can be increased by using a large vapor cell.  

Figure~\ref{fig:QNL} shows the fundamental quantum noise limit of OQS for operation using K spins, calculated using Eq.~(\ref{eq:QNL}) with $\gamma =7\times 10^9~\text{Hz/T}$, $n = 7 \times10^{13}~\text{cm}^{-3}$, $V = 100~\text{cm}^3$, and $T_2 = 3.5~\text{ms}$.  It indicates that the OQS field noise of 10~aT/$\sqrt{\text{Hz}}$ can be achievable at the optimal condition. The K atoms (natural abundance mixture of two stable isotopes K-39 and K-41 both having the same nuclear spin 3/2) are selected owing to their lower spin-destruction cross section between atoms~\cite{PhysRevLett.95.063004} (10 times lower than that of Rb atoms). The optimal value of $R_{pr}\times \text{OD}$ has to be quite large. As the $R_{pr}\times \text{OD}$ is proportional to the alkali-metal density, the length of the vapor cell along the probe beam direction, and the probe laser power~\cite{PhysRevLett.95.063004}, the large optimal value can be achieved by optimally increasing the three parameters. The excessive increase of the alkali-metal density will shorten $T_2$; hence, the density should increase until alkali-metal spin-destruction collisions start to dominate the spin-destruction rate. Increasing the probe beam path length, which does not affect $T_2$, can be implemented by using a vapor cell that has a long dimension along the probe beam direction; however, this method is not practical if a path length of greater than 20~cm is required. The probe laser power can be increased as long as it does increase much $1/T_2$~\cite{PhysRevLett.95.063004}. 

\subsection{Intrinsic Magnetic Johnson Noise of LC Circuit}

The magnetic Johnson noise of the circuit $\delta B_{J}$ is the combined thermal noise from the input gradiometer coil, the output coil, and the wire leads between the input and output coils (Fig.~\ref{fig:config}). Due to current fluctuations arising from the voltage Johnson noise of the LC circuit, $ \delta I_{J}=\sqrt{4k_BT R}/Z=Q\sqrt{4k_BTR}/\omega (L_{\text{in}}+L_{\text{out}})$ with $Z$ being the impedance of the resonant LC circuit, the $\delta B_{J}$ at the location of the OQS vapor cell is given by
\begin{align}
   \delta B_{J}= &\Big[{\omega (L_{\text{in}}+L_{\text{out}}) r_{\text{out}} }[1+(d/r_{\text{out}})^2]^{3/2}\Big]^{-1}\notag\\\times & N_{\text{out}}Q\sqrt{4k_BT\rho \Big[\frac{4(l_{\text{in}} + r_{\text{in}})N_{\text{in}}}{A_{\text{in}}}+\frac{4\pi r_{\text{out}}N_{\text{out}}}{A_{\text{out}}}+\frac{2l_{\text{lead}}}{ A_{\text{lead}}}\Big]},
   \label{eq:delBJ}
\end{align}
where $k_B$ is the Boltzmann's constant; $T$ is the absolute temperature of the LC circuit; $l_{\text{lead}}$ is the length of the wire leads; $\rho$ and $A_{\text{in,out,lead}}$ are the the resistivity and the total cross section area of the wires of the input gradiometer coil, output coil, and wire leads, respectively. This indicates that the $\delta B_{J}$ reduces at lower temperature. 

\subsection{Backaction OQS noise}
We investigate backaction of the OQS, which injects the spin noise arising from fluctuating K spins in the OQS cell into the LC circuit through the output coil. The backaction OQS noise becomes significant when using the LC circuit cooled to a few K due to its low thermal noise. The OQS spin noise can be estimated in the assumption of the standard quantum limit at the magnetic resonance: 
\begin{equation}
  \delta B_{SN}=\frac{\mu_0\mu_B \sqrt{N_s T_2}}{4 \pi d^3  }L(\nu-\nu_0, \Delta \nu),
  \label{eq:SpinNoise}
\end{equation} 
where $\mu_B$ is the Bohr magneton, $d$ is the distance from the center of the cell, $N_s$ is the number of the K spins in the cell. Here we consider the K cell as a magnetic dipole with the fluctuating magnetic moment of $(1/2)\mu_B \sqrt{N_s T_2}$ for simplification.  We assume that the spin noise scales as $\sqrt{N_s}$ in the standard quantum limit and it has a Lorentzian profile $L(\nu-\nu_0, \Delta \nu)$ near the magnetic resonance \cite{ShahSpinNoise}. This profile is normalized to 1 at the maximum and has the relation of $\Delta \nu=1/(2 \pi T_2)$. For the K cell that can reach  $\delta B_{\text{QNL}}=10$~aT/$\sqrt{\text{Hz}}$ at the optimal condition, discussed in Sec.~\ref{subsec:QNL}, with $V=100~\text{cm}^3$ and $n=7\times 10^{13}$ cm$^{-3}$, the spin noise is estimated to be $\delta B_{SN}=0.009$~aT/$\sqrt{\text{Hz}}$ at $d=8$~cm, as an example.

The magnetic flux of $\delta B_{SN}$ through the two-loop circular output coil is 
\begin{align}
  \delta \Phi_{SN} = &  \int 2N_{\text{out}} \delta B_{SN} dA \notag \\
  & = \frac{ \mu_0\mu_B \sqrt{N_s T_2}  N_{\text{out}}}{r_{\text{out}}[1+(d/r_{\text{out}})^2]^{3/2}}L(\nu-\nu_0, \delta \nu),
  \label{eq:backaction}
\end{align} 
which injects the OQS spin noise into the LC circuit through the relation $\delta V_{SN}=\delta \Phi_{SN} \omega.$ The $\delta V_{SN}$ should be compared with the thermal noise inside the LC circuit in order to check the significance of the backaction OQS noise.

\subsection{Possible Magnetic Background Noise}
The main experimental challenge is to detect the extremely weak magnetic axion signal above magnetic background noises. The LC circuit needs to be shielded from external electromagnetic fields. The simplest shielding method is placing the elements of the LC circuit into a copper shield (but not any ferromagnetic shield, since the circuit is partially exposed to the strong magnetic field). We estimate that the shielding factor of $10^3$ ($10^6$) can be achieved with a copper shield thickness $x = 0.5$~mm (1~mm) based on the $e^{-x/\delta}$ shielding law with the copper’s skin depth $\delta = 65~\mu$m at 1~MHz. The most sensitive part of the LC circuit to the electromagnetic interference is the large input gradiometer coil. The output coil and the OQS sensor head are not exposed to the strong field, therefore various options exist for shielding including $\mu$-metal shield and ferrite shield that has the lowest possible magnetic background noise due to its extremely small electrical conductivity. Since the non-ferromagnetic conductive copper shield generates magnetic background noise and can impose limitations on the axion detector sensitivity, we consider the question of copper shield thermal noise in detail in this subsection.   
 
\subsubsection{Copper Shield Thermal Noise: Johnson Noise and  Black-Body Radiation Noise}
The thermal noise of the conductive copper shield can potentially have two components: the Johnson noise and the black-body radiation noise. Here, we show that actually at frequencies below $\sim$MHz, the black-body radiation noise is identical to the Johnson noise for the copper surface, and thus its contribution is included via the Johnson noise. While one might try to use Plank's law to obtain the energy density for black-body radiation
\begin{equation}
 \epsilon(\omega)\propto \frac{\omega^3}{e^{\hbar \omega/(k_B T)}-1},  
\end{equation}
at the frequency range the assumptions which were used for the derivation are no longer true, and also this limit is not well understood~\cite{Guo}. Physically thermal black-body radiation arises from thermal motion of atoms~\cite{Guo2016}, which leads to random current densities~\cite{RandomCurrent}. By using the expression for the photon energy Bose-Einstein distribution  
\begin{equation}
  \Theta(\omega,T)=    \frac{\hbar\omega}{e^{\hbar \omega/(k_B T)}-1},
\end{equation}
that in the frequency limit of $\hbar\omega <<k_B T $ can be simplified to 
\begin{equation}
  \Theta(\omega,T)=   k_B T,
\end{equation}
and substituting the dielectric constant from the Drude model (see for example Ref.~\cite{Basu})
\begin{equation}
    \epsilon(\omega)=\epsilon_\infty-\frac{\sigma_0/\tau}{\epsilon_0(\omega^2+i\omega/\tau)},
\end{equation}
where $\varepsilon_\infty$ is the $\varepsilon(\omega)$ at very high frequency, $\sigma_0$ is the electrical conductivity at zero frequency, $\varepsilon_0$ is the vacuum permittivity, and  $\tau$ is electron collision time, these random current densities can be written as
\begin{equation}
    <j_m(r,\nu) j_n(r',\nu')>=4 \sigma k_B T \delta_{mn}\delta(r-r')\delta(\nu-\nu').
    \label{currentnoise}
\end{equation}
Here $\sigma$ is the electrical conductivity and $r$ ($r'$) is the $x$, $y$, $z$ ($x'$, $y'$, $z'$) vector.
Note that we replaced $\delta(\omega-\omega')$ in the original expression of Ref.~\cite{RandomCurrent} with $\delta(\nu-\nu')/2\pi$ and multiplied it by a factor of 2 for folding the negative spectrum on to the positive one.

It can be shown that the expression in Eq.~(\ref{currentnoise}) leads to a correct Johnson noise voltage. The black-body radiation current in a short copper can be found by integrating the fluctuating random current densities over the cross-section area $A$ of the copper. This gives
\begin{equation}
     I_{BB}^2(z,\nu)=4 \sigma k_B T \delta(z-z')\delta(\nu-\nu').
\end{equation}
The black-body radiation voltage is then $dV_{BB}=I dR=I dz/\sigma A$,
and by two-time integration of the squared voltage over the length of the conductor $l$ along  $z$, 
\begin{equation}
    V_{BB}^2=4 k_B T R  \delta(\nu-\nu').
\end{equation}
Here $R=l/\sigma A$. 
The delta function means un-correlated noise in the frequency domain, therefore double integration over some small frequency interval $\Delta\nu$ leads to the correct expression for Johnson noise $V_{JN}$
\begin{equation}
    V_{BB}=\sqrt{4 k_B T R  \Delta\nu}=V_{JN}.
\end{equation}

The magnetic noise of a small copper disk of radius $a$ at a distance $z$ much larger than $a$ can be easily calculated analytically from Eq.~(\ref{currentnoise}) and the Biot-Savart law:
\begin{equation}
    B=\int \frac{\mu_0}{4\pi}\frac{j\times r}{r^3}dxdydz.
\end{equation}
Since $B$ is the random function, we calculate $B^2$, actually $B_z^2$, the component normal to the surface of the copper disk, which is proportional to 
\begin{equation}
    <(j_x y-j_y x)^2>=<j_x,j_x>y^2+<j_y,j_y>x^2=j^2_0(x^2+y^2),
\end{equation}
where $j^2_0=4 \sigma k_B T \delta(r-r')\delta(\nu-\nu')$.
This can be integrated over the volume of the disk with the thickness $h$ as well as over frequency similar to such integration in case of the copper voltage noise to give
\begin{equation}
     B_z=\frac{\mu_0}{\sqrt{8\pi}}\sqrt{\sigma k_B T h}\frac{a^2}{z^3},
\end{equation}
which in the limit of $a\ll z$ is the exact expression for the Johnson noise of a small disk~\cite{RomalisNoisePaper}.
In case of an arbitrary copper shield, since we showed that the black-body radiation noise is identical to the Johnson noise, it is possible just to use Johnson noise calculations in order to estimate the thermal noise of the copper shield. 

There is still a question of noise arising from dielectric materials, for example, glass material of the OQS vapor cell. For the noise from a dielectric material, the random current density in Eq.~(\ref{currentnoise}) can be obtained by replacing $\sigma$ with $\omega \epsilon_0 \epsilon''$, where $\epsilon''$ is the imaginary part of the complex dielectric permittivity. The noise ratio between copper and dielectric materials for the same geometry will be 
\begin{equation}
    \sqrt{\frac{\sigma \delta_S}{\omega \epsilon_0 \epsilon'' h}}.
\end{equation}
If we compare a copper disk with $\sigma=5.96\times10^7$~S/m and the skin depth $\delta_S=7\times 10^{-5}$ m with a disk made of dielectric material with a typical $\epsilon''=10^{-3}$ and the thickness $h=1$~cm for $\omega=2\pi\times 10^6$~Hz, the noise ratio is $2.7\times 10^6$, meaning that the noise due to dielectric material losses is expected much lower and can be neglected. 

\subsubsection{Magnetic Johnson Noise of Copper Shield}
The magnetic field noise due to Johnson current noise in the copper shield can be found using the method in Ref.~\cite{doi:10.1063/5.0029998}. The noise for specific copper shield configuration can be obtained by scaling the results in Ref.~\cite{doi:10.1063/5.0029998}. In particular, it has been shown that the noise at considered frequency scales, where the inductance of the current path dominates over the resistance of the path, in the following way: 
\begin{equation}
    \delta B=k  T^{1/2}\sigma^{-1/2}\omega^{-3/4} h^{-2},
\end{equation}
where $k$ is a coefficient of proportionality dependent on other parameters such as temperature and shield geometry.

In Ref.~\cite{doi:10.1063/5.0029998} it has been shown that at a 4.6 cm distance and room temperature, the thermal noise of copper shield with the thickness $h=0.2$~mm for $\omega=2\pi\times 10^5$~Hz, was measured to be 220~aT$\sqrt{\text{Hz}}$ and the shield Johnson noise was modeled to be 210~aT$\sqrt{\text{Hz}}$. This result confirms our claim that the thermal noise of the copper shield at frequencies below $\sim$~MHz can be estimated by only the Johnson noise of the copper shield. The noise value can be scaled to a 50~cm distance (the approximate average distance from the copper shield to the coils in our experiment) and 1~MHz frequency, giving 0.3~aT/$\sqrt{\text{Hz}}$. Further reduction in the noise can be achieved by cooling the copper material. If nitrogen cooling is used (77~K), the copper resistivity can be reduced by 10 times (slightly depending on purity of copper material). Taking into consideration of the temperature reduction factor, the noise at 77~K can be reduced by 6 times to 0.05~aT/$\sqrt{\text{Hz}}$. Cooling the copper material to 4 K can reduce the noise another factor of 40 for high-purity copper, reaching 0.001~aT/$\sqrt{\text{Hz}}$.

\section{Prototype Axion Detector Development}\label{sec:exp}
\begin{figure}[t!]
\centering
\includegraphics[width=3.4in]{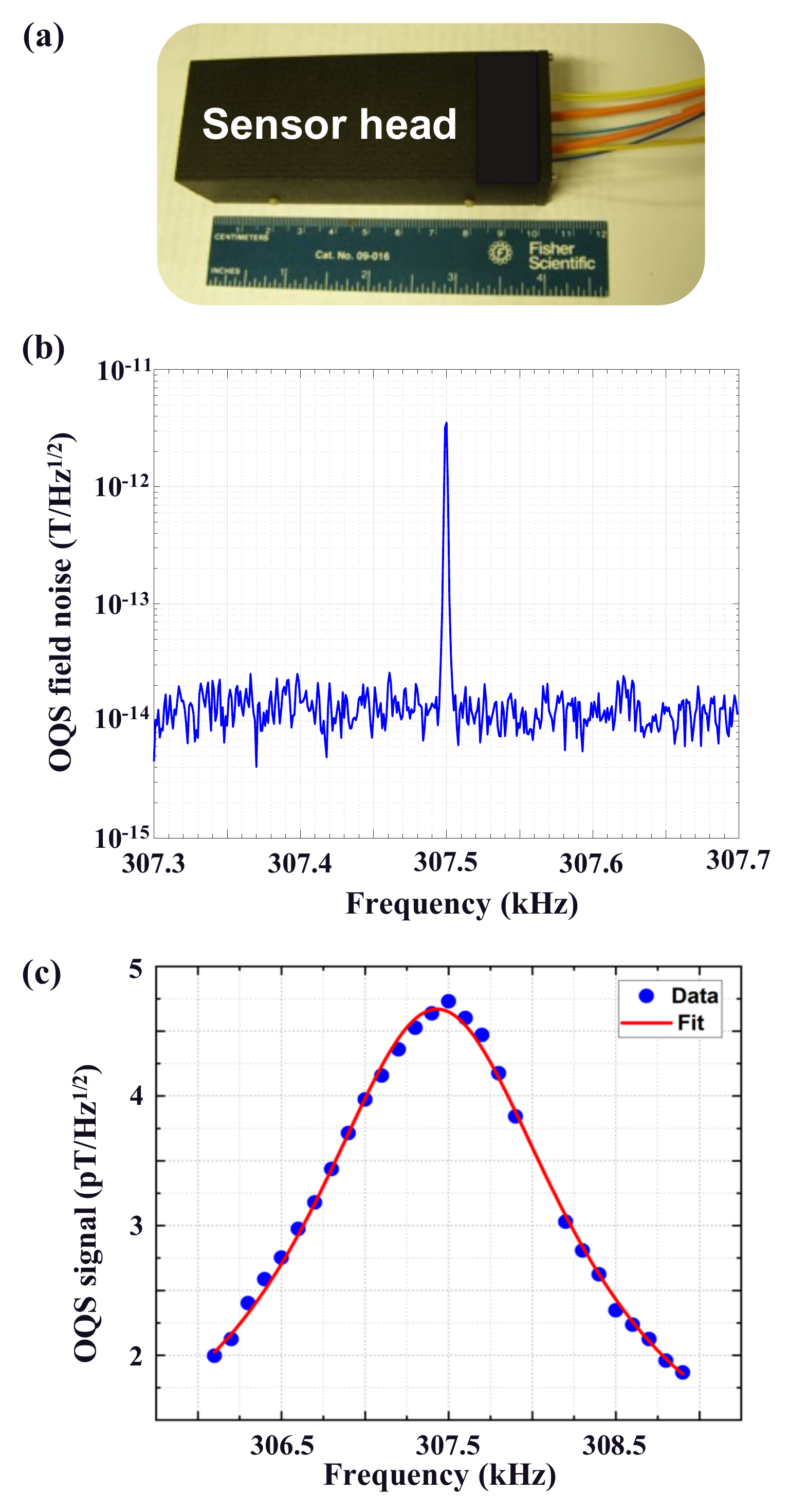}
\caption{(a) Photograph of the cm-scale commercial OQS sensor head, (b) the magnetic field noise and (c) the frequency response of the OQS. In (b), the peak at 307.5~kHz is the known applied calibration magnetic field. The magnetic field noise of the OQS is measured to be 10~fT/$\sqrt{\text{Hz}}$. In (c), the data was collected by scanning applied sine field of constant amplitude between 306 and 309~kHz; the solid curve indicates a Lorentzian fit, giving the OQS bandwidth of 1.8~kHz.}
\label{fig:OPM}
\end{figure}

To perform the first tests of an OQS in axion detection with an LC circuit, we have developed and constructed an optimized room temperature design for an existing solenoid magnet at Los Alamos
National Laboratory (LANL), with a commercial OQS.
The magnet is a superconducting solenoid with a warm bore of 1-m diameter and 3-m length in the center, which can produce a static field $B_0 = 2$~T. We employed a commercial Twinleaf OQS containing a $5\times5\times5~\text{mm}^3$ Rb-87 vapor cell, whose sensor head is shown in Fig.~\ref{fig:OPM}(a). The measured magnetic field noise of the OQS around 300~kHz (the target frequency range in this prototype experiment described below) was 10~fT/$\sqrt{\text{Hz}}$. As an example, the OQS field noise around 307.5~kHz is shown in Fig.~\ref{fig:OPM}(b). The peak at 307.5~kHz is a known applied calibration magnetic field to convert the OQS output voltages into magnetic fields. We also investigated a frequency response of the OQS by applying a sinusoidally varying magnetic field at different frequencies around 307.5~kHz, as indicated in Fig.~\ref{fig:OPM}(c) (blue dot points). The data was fit to a Lorentzian function, $f(\nu)=a_0+a_1/[4(\nu-\nu_0)^2+\Delta\nu^2]$ with $\nu$ being the frequency of the applied field. The fit (solid red curve) gives the bandwidth of the OQS of $\Delta\nu_\text{OQS}=1.8~$kHz. 

\begin{figure}[t!]
\centering
\includegraphics[width=3.4in]{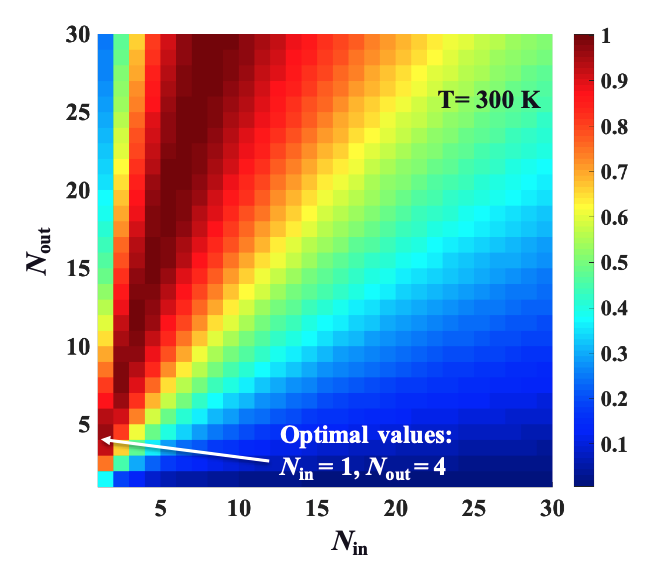}
\caption{Optimization of the number of turns of the input gradiometer and output coils to maximize the SNR of the prototype axion detector. The SNR values are normalized. }
\label{fig:optimization}
\end{figure}

To accommodate the sizes of the magnet bore and the OQS, we selected the following experimental dimensions for the axion detector: $l_{\text{in}}=1.0$~m, $r_{\text{in}}=0.3$~m, $r_{\text{out}}=2.9$~cm, and $d=2.0$~cm. The $l_{\text{lead}}=9.0$~m was chosen in order to locate the OQS at a position where the residual magnetic field of the magnet is sufficiently suppressed to the level of the Earth's magnetic field. To reduce the AC resistance of the LC circuit closer to the DC resistance value, the circuit was made from Litz wires of multiple strands of thin copper wire with a 0.06~mm diameter, recommended at our target frequency range~\cite{Litz}. Because the total length of the LC circuit is dominated by the input coil and the wire leads, their wire diameter was chosen to be $b_{\text{in}}=5.2$~mm in order to sufficiently reduce their magnetic Johnson noise. On the other hand, the wire diameter of the output coil was reduced to $b_{\text{out}}=1.3$~mm because of the small output coil's diameter.  

\begin{figure}[t!]
\centering
\includegraphics[width=3.6in]{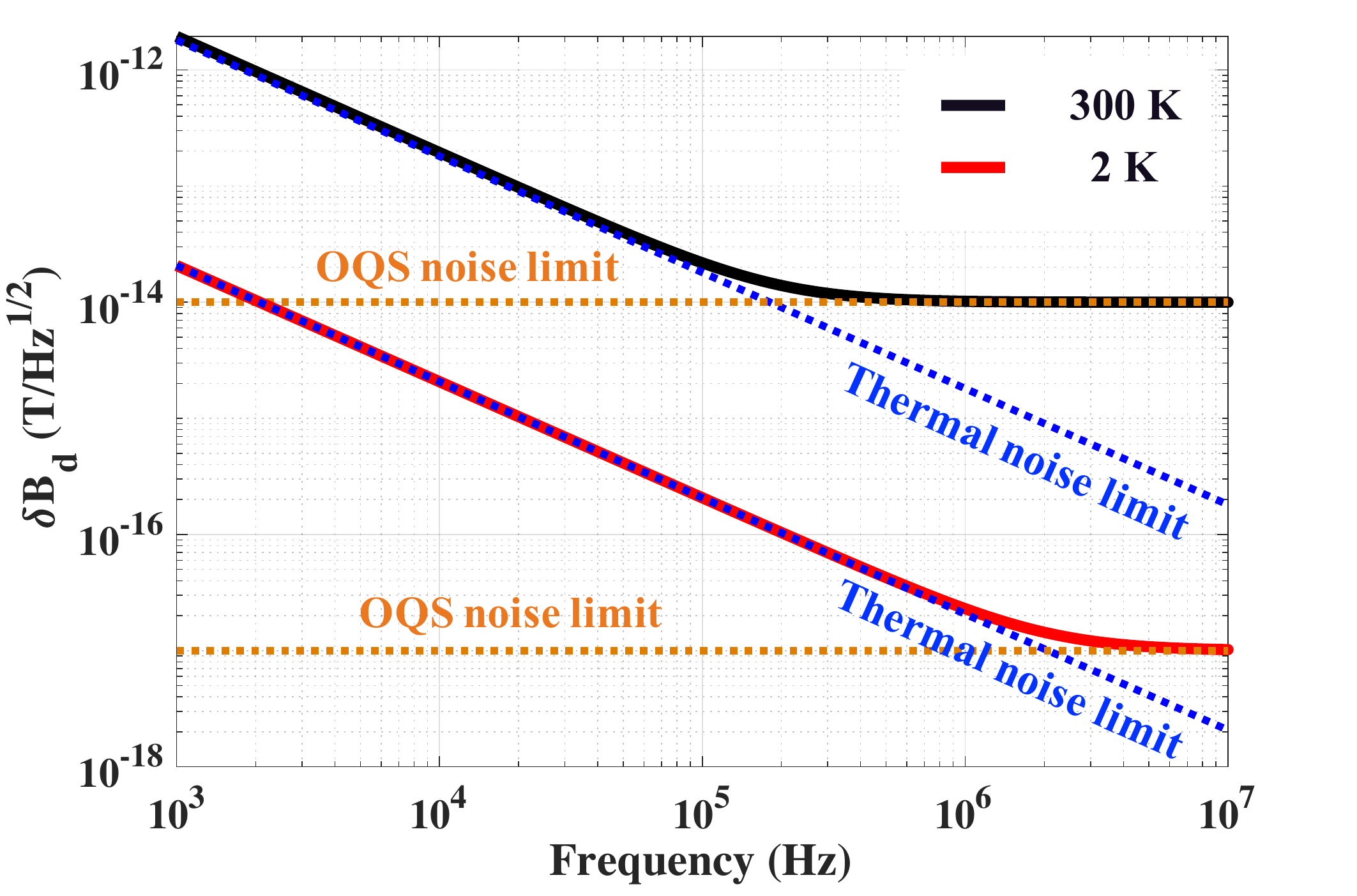}
\caption{Estimated total magnetic noise of the prototype axion detector with the optimized LC circuit at 300~K and the commercial OQS unit (black solid curve), and the advanced axion detector with the optimized superconducting LC circuit at ~2~K and the quantum-limited OQS (red solid curve). The noise below $\sim$100~kHz and $\sim$1~MHz is dominated by the thermal noise of the LC circuit at 300~K and 2~K, respectively, limiting the axion detector sensitivities.}
\label{fig:DelBd}
\end{figure}

We optimized $N_{\text{in}}$ and $N_{\text{out}}$ to maximize the signal-to-noise ratio (SNR) of the axion detector, $|B_d|/\delta B_d$, at 300~kHz using Eqs.~(\ref{eq:bd_out2}), (\ref{eq:delBd}), and (\ref{eq:delBJ}), and $\delta B_{\text{OQS}}=10$~fT/$\sqrt{\text{Hz}}$. We estimated the inductance of the input gradiometer coil~\cite{PhysRevLett.112.131301},
\begin{align}
L_{\text{in}}\approx \frac{2}{\pi} N_{\text{in}}^2l_{\text{in}} \text{ln}(r_{\text{in}}/0.5b_{\text{in}})=N_{\text{in}}^2\times3.8~\mu \text{H},
\label{eq:L_in}
\end{align}
and the inductance of the output coil~\cite{PhysRevLett.112.131301},
\begin{align}
L_{\text{out}} \approx 2r_{\text{out}} N_{\text{out}}^2 \Big[\text{ln}\Big(\frac{8r_{\text{out}}}{0.5b_{\text{out}}}\Big)-2\Big]= N_{\text{out}}^2\times0.3~\mu \text{H},
\label{eq:L_out}
\end{align}
where we neglected the mutual inductance between the two loops for simplicity. Figure~\ref{fig:optimization} shows the calculated, normalized 2D distribution of the SNR values of the axion detector as a function of $N_{\text{in}}$ and  $N_{\text{out}}$. This calculation implies that various optimal values exist in the red region; however, fewer number of turns are better according to our investigations that the noise of coils approached the theoretical value in Eq.~(\ref{eq:delBJ}) as the number of turns was reduced~\cite{SAVUKOV2009188}. Thus, we selected $N_{\text{in}} = 1$ and $N_{\text{out}}= 4$ as optimal values, leading to the input gradiometer consisting of two series-configured one-turn $1.0~\text{m}\times0.3~\text{m}$ pickup loops. Figure~\ref{fig:DelBd} shows the estimated total magnetic noise of the optimized prototype axion detector as a function of the frequency (black solid curve), indicating that the axion detector loses sensitivity at the frequency range below $\sim$100~kHz due to the thermal noise of the room temperature LC circuit (blue dotted curve). Above $\sim$100~kHz, our experimental sensitivity to the axion is limited by the OQS field noise (orange dotted curve). Based on Fig.~\ref{fig:DelBd}, our experiment targeted frequencies around 300~kHz. 

\begin{figure}[t!]
\centering
\includegraphics[width=3.4in]{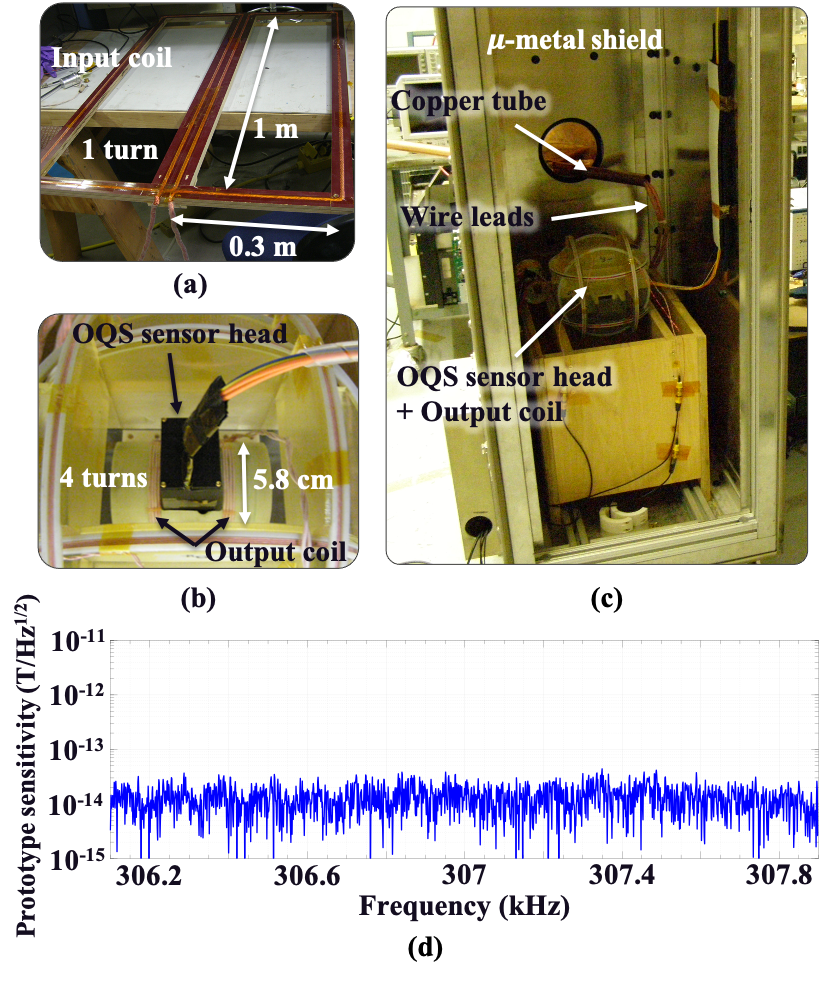}
\caption{The Photographs of the prototype axion detector: (a) the optimized first-order planar input gradiometer coil, (b) the combination of the optimized circular output coil and the commercial OQS sensor head, (c) the $\mu$-metal enclosure that houses the combination shown in (b) (the $\mu$-metal lid is opened for illustration purposes only),  and (d) its sensitivity when the magnet is cooled to 4~K and the prototype is tuned at 307~kHz. }
\label{fig:setup}
\end{figure}

We have built a prototype axion detector with the selected experimental dimensions and the optimized experimental parameters, as shown in Fig.~\ref{fig:setup}(a)--(c). The wire leads were shielded by copper tube and the assembly of the output coil and the OQS sensor head was shielded by a $\mu$-metal enclosure in order to minimize interference from ambient magnetic signals. On the other hand, the input gradiometer coil was shielded by the magnet shield composed of an open iron rectangular enclosure located outside the magnet bore and additional copper mesh and Faraday cage to cover the openings. When the magnet is cooled to 4~K, the superconducting solenoid coil of the magnet could also add additional shielding for the input gradiometer coil. The OQS output was connected to a 24-bit data acquisition system (NI PXIe-4480) and recorded at a sampling rate of 1~MHz using a home-built LabVIEW program. 

First, we tuned the prototype axion detector around 307~kHz. This was achieved by both tuning the OQS by applying its corresponding bias magnetic field of $43.9~\mu\text{T}$ and tuning the LC circuit by using mica capacitors of 19~nF. The quality factor of the LC circuit was measured to be $Q=43$, indicating its bandwidth of 7.1~kHz through the relation of $\Delta \nu_{\text{LC}}=\nu/Q$. Since $\Delta \nu_{\text{LC}} > \Delta \nu_{\text{OQS}}$, the bandwidth of the prototype is determined by $\Delta \nu_{\text{OQS}}=1.8~\text{kHz}$. Before the magnet was cooled, the sensitivity of the prototype was measured to be 50~fT/$\sqrt{\text{Hz}}$, 5 times larger than the estimated values shown in Fig.~\ref{fig:DelBd}. We estimate that the worse sensitivity was due to the magnetic Johnson noise from the solenoid coil of the magnet at room temperature, but this noise decreased significantly when the solenoid coil became superconducting.

With the magnet cooled to 4~K where the solenoid coil becomes superconducting, we obtained background data with the prototype (i.e., without the magnet energized), shown in Fig.~\ref{fig:setup}(d). The experimental sensitivity of the prototype was measured to be around 10~fT/$\sqrt{\text{Hz}}$, demonstrating that with other ambient noises sufficiently suppressed, the prototype sensitivity to the axion dark matter is determined by the OQS noise. This means that the OQS noise reduction is the key to success. 

\section{Sensitivity Estimate of Axion detection}\label{sec:sens}
In principle, the SNR of the prototype axion detector can be increased with long data integration. The total field noise of the prototype with a data integration time, $t_{\text{int}}$, is given by
\begin{align}
\delta B_{d}^{\text{int}}=\delta B_{d}\times(t_ct_{\text{int}})^{-1/4},
\label{eq:delbd_int}
\end{align}
where $t_c=(0.16~\text{s})\times(\text{MHz}/\nu)$ is the axion signal coherence time for the isothermal halo model~\cite{PhysRevLett.112.131301, PhysRevX.4.021030}. For example, the $\delta B_{d}$ can be reduced by a factor of 11 and 8 with a 7-hours integration time at 300~kHz and 1~MHz, respectively. The axion signal coherence time limits the experimental noise reduction with long data integration. 

\begin{figure}[t!]
\centering
\includegraphics[width=3.4in]{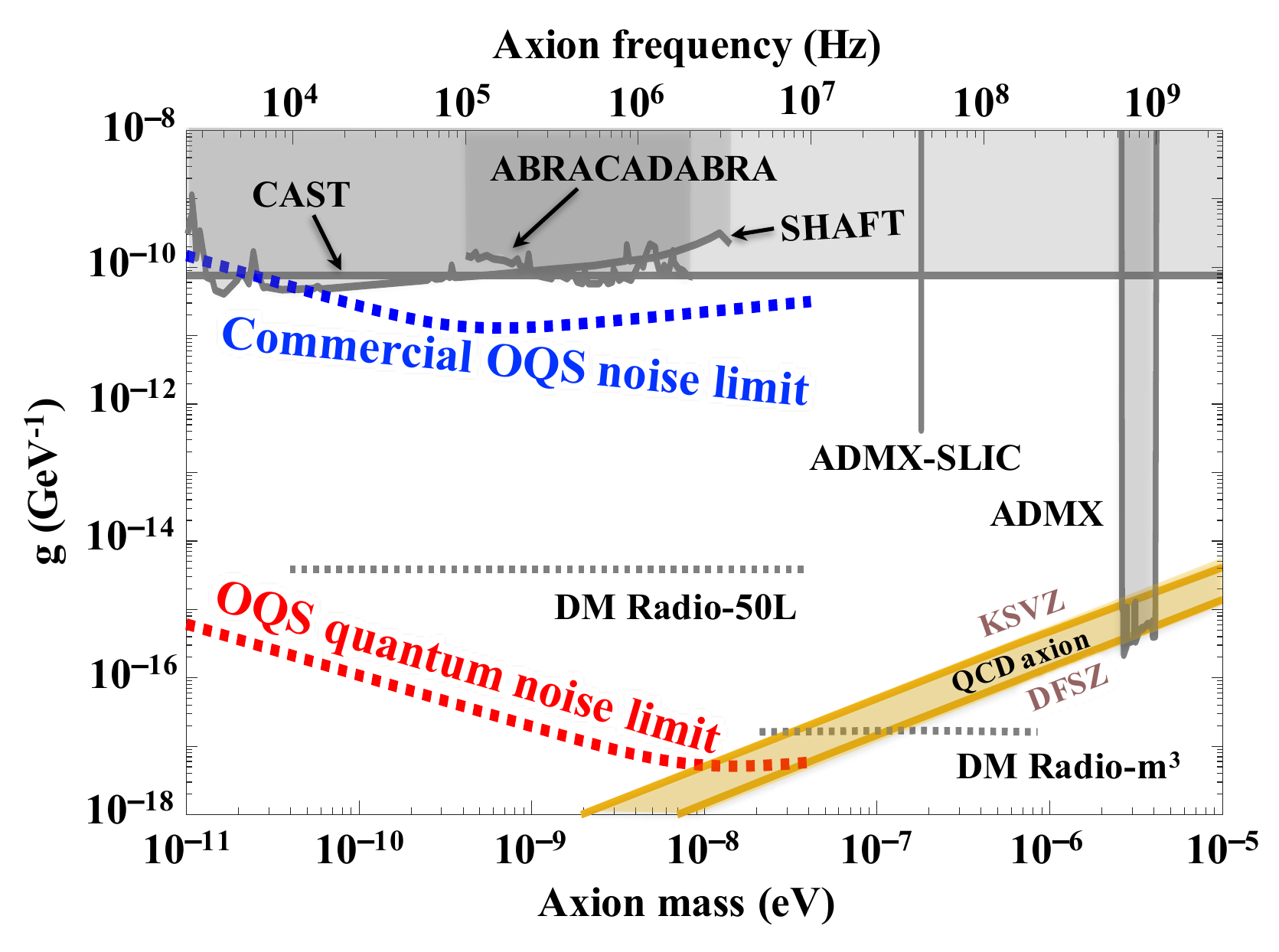}
\caption{Estimated sensitivity of our prototype axion detector (blue dotted curve) to the axion-photon coupling $g$ with $t_{\text{int}}=7~\text{hours}$ and 95\% confidence level to the axion mass range from $10^{-11}$ to $4\times 10^{-8}$ eV. In similar range, the constraints were set by the CAST~\cite{CAST}, ABRACADABRA~\cite{PhysRevLett.127.081801}, and SHAFT~\cite{SHAFT} experiments (gray solid curves). Estimated sensitivities of the DM-Radio experiments (gray dotted curves) are also shown~\cite{PhysRevD.106.112003}. The yellow band encompasses various QCD axion models, including the benchmark KSVZ~\cite{PhysRevLett.43.103,SHIFMAN1980493} and DFSZ~\cite{DINE1981199,osti_7063072} models, predicting the QCD axion coupling. Furthermore, the constraints set by the ADMX~\cite{PhysRevLett.120.151301,PhysRevLett.124.101303,PhysRevLett.127.261803} and ADMX-SLIC~\cite{PhysRevLett.124.241101} experiments are shown. The red dotted curve shows the estimated sensitivity of our proposed axion detector optimized to an quantum-limited OQS with 10~aT/$\sqrt{\text{Hz}}$ field noise and a superconducting LC circuit with $t_{\text{int}}=15$~s.}
\label{fig:sen}
\end{figure}

Based on Eqs.~(\ref{eq:bd_out2}), (\ref{eq:delBd}), (\ref{eq:delBJ}), and (\ref{eq:delbd_int}), we can estimate the sensitivity of the prototype axion detector to the axion-photon coupling $g$,
\begin{align}
g &= \text{S}\frac{r_{\text{out}} [1+(d/r_{\text{out}})^2]^{3/2} (L_{\text{in}}+L_{\text{out}})\delta B_{d}^{\text{int}} }{2QN_{\text{in}} N_{\text{out}}V_{\text{in}}\sqrt{2\rho_{DM}}B_0}\notag\\
&= \text{S}\Big(\frac{\delta B_{d}^{\text{int}}}{10^{-15}~\text{T}}\Big)\Big(\frac{\text{GeV/cm}^3}{\rho_{DM}}\Big)^{\frac{1}{2}}\Big(\frac{10^3}{Q}\Big)\Big(\frac{L}{\mu \text{H}}\Big)\Big(\frac{\text{T}}{B_0}\Big)\Big(\frac{1}{N_{\text{out}}}\Big)\notag\\&\times \Big(\frac{1}{N_{\text{in}}}\Big)\Big(\frac{r_{\text{out}}[1+(d/r_{\text{out}})^2]^{\frac{3}{2}}}{\text{cm}}\Big)\Big(\frac{\text{m}^3}{V_{\text{in}}}\Big)(2\times10^{-16}~\text{GeV}^{-1}),
\label{eq:g_sen}
\end{align}
where S is the SNR, taken as 2 ($2\sigma$ or 95\% confidence level; note that once the axion signal is discovered, the measurement can be repeated many times to verify the discovery with higher confidence level), and $L=L_{\text{in}}+L_{\text{out}}$ is the total inductance of the LC circuit. Fig.~\ref{fig:sen} shows our estimated sensitivity of the prototype (blue dotted curve) from the background data obtained with 7-hours integration at each observation frequency.  The upper end of our search range, around 10~MHz, is limited by the combination of inductance and stray capacitance of the chosen specific configuration of the LC circuit. The size of the coil can be reduced in principle to extend the search to higher frequencies, but this would require redesigning the coil and will also reduce the axion flux and hence sensitivity. The sensitivity loss at masses larger than 10$^{-9}$~eV is due to the noise reduction limitation from the axion
signal coherence time, described above. Our prototype sensitivity could improve the current constraints (gray solid curves) set by the CERN Axion Solar Telescope (CAST)~\cite{CAST}, Broadband/Resonant Approach to Cosmic Axion Detection with an Amplifying B-field Ring Apparatus (ABRACADABRA)~\cite{PhysRevLett.127.081801}, and Search for Halo Axions with Ferromagnetic Toroids (SHAFT)~\cite{SHAFT} experiments on an axion mass range between 10$^{-10}$~eV and 10$^{-7}$~eV, corresponding to the frequency range between 10~kHz and 10~MHz, in particular by 1 order of magnitude at masses of around 10$^{-9}$~eV. For comparison, projected sensitivities of the DM-Radio experiment are shown with two gray dotted lines~\cite{PhysRevD.106.112003}. The yellow band indicates a broad range of the axion-photon
coupling $g$ for the QCD axion predicted by various axion models. As benchmark examples, the KSVZ~\cite{PhysRevLett.43.103,SHIFMAN1980493}  and DFSZ~\cite{DINE1981199,osti_7063072} axion models are included. The constraints set by the ADMX~\cite{PhysRevLett.120.151301,PhysRevLett.124.101303,PhysRevLett.127.261803} and ADMX-SLIC (Superconducting LC Circuit Investigating Cold Axions)~\cite{PhysRevLett.124.241101} experiments are also shown. Our experiment will be able to probe the axion dark matter on the mass range between 10$^{-11}$~eV and 10$^{-7}$~eV. 

\section{Potential Improvement of Axion Detector Sensitivity}\label{sec:imp}

Figure~\ref{fig:DelBd} shows the magnetic Johnson noise of optimized LC circuits at 300~K (top blue dotted curve) and 2~K (bottom blue dotted curve),
and the OQS noise limit of the commercial unit (top orange dotted curve), and a quantum-limited OQS (bottom range dotted curve). It is clear that to take advantage of the quantum-limited OQS in axion detection, the detector thermal noise must be reduced by cooling the LC circuit. In this regime, the backaction noise from the OQS and the thermal noise from the surrounding copper shield must also be considered, as discussed in Section~\ref{sec:noise}. An optimized design of an axion detector with a quantum-limited OQS is considered in this section.

\subsection{Axion Detector with Quantum-limited OQS}
Improvement over the prototype design can be achieved with a quantum-limited OQS with $\delta B_{\text{OQS}}=10$~aT/$\sqrt{\text{Hz}}$. To take full advantage of a quantum-limited OQS, we consider the LC circuit made of a pure superconducting wire [e.g., 3-mil niobium (Nb) wire] and cooled to 4~K and below, which can considerably reduce the magnetic Johnson noise of the LC circuit. We investigate the sensitivity of an improved axion detector with $l_{\text{in}}=2.9$~m and $r_{\text{in}}=0.45$~m, more closely matching the dimension of the bore of the magnet to maximize the axion-induced magnetic flux through the input gradiometer coil. Considering the K cell volume of $V = 100~\text{cm}^3$, we selected $r_{\text{out}}=5.0$~cm and $d=8.0$~cm to allow space between the cooled output coil and the K cell. We also selected $l_{\text{lead}}=4$~m where the residual magnetic field of the magnet is reduced to the level of the Earth's magnetic field. 

The inductance of the superconducting input and output coils was estimated using a three-dimensional inductance extraction program in superconducting structures~\cite{FastHenry}: $L_{\text{in}}=N_{\text{in}}^2\times26.0~\mu \text{H}$ and $L_{\text{out}}=N_{\text{out}}^2\times0.4~\mu \text{H}$. The parasitic inductance of the twisted-pair leads of 3-mil superconducting Nb wires was measured to be 2.3~nH/cm~\cite{NbInductance}, giving the negligible inductance of the wire leads of $0.9~\mu$H. While a pure superconducting wire has zero electrical resistance at low frequencies, the resistance at high frequencies is extremely small but non-zero~\cite{Superconducting}. The high-frequency resistance $R_{Nb}$ of superconducting Nb at around 2~K is on the order of n$\Omega$~\cite{Superconducting}. This $R_{Nb}=1~\text{n}\Omega$ resistance can achieve $Q\approx10^{11}$, however we will detune the superconducting LC circuit to reach $Q_{\text{eff}}
\approx10^{6}$ in order to have a reasonable axion detector bandwidth. Based on Eq.~(\ref{eq:delBJ}), the corresponding thermal magnetic Johnson noise at the location of the OQS vapor cell is given by
\begin{align}
   \delta B_{J}= \frac{N_{\text{out}}Q_{\text{eff}}\sqrt{4k_BTR_{Nb}}}{\omega (L_{\text{in}}+L_{\text{out}}) r_{\text{out}}[1+(d/r_{\text{out}})^2]^{3/2}}.
   \label{eq:delBJ_Nb}
\end{align}

\begin{figure}[t!]
\centering
\includegraphics[width=3.4in]{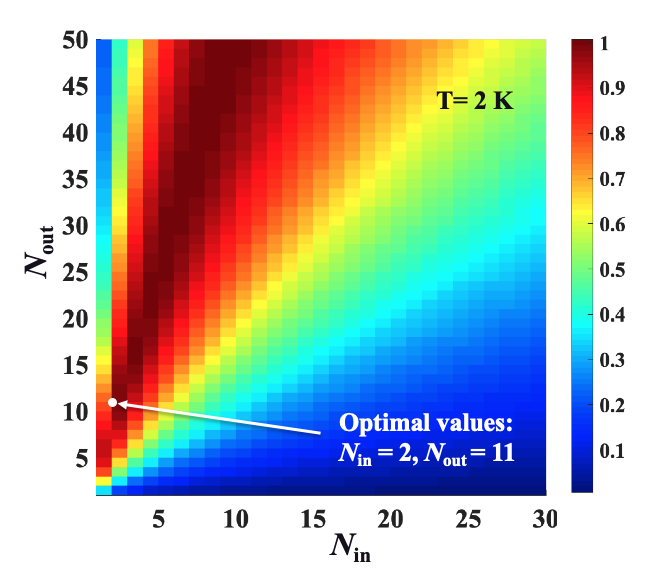}
\caption{Optimization of the number of turns of the input gradiometer and output coils made of 3-mil Nb wire at 2~K to maximize the SNR of the axion detector with a quantum-limited OQS. The SNR values are normalized. }
\label{fig:opt_Nb}
\end{figure}

For the axion detector with the quantum-limited OQS, we re-optimized $N_{\text{in}}$ and $N_{\text{out}}$ to maximize the SNR of the detector, $|B_d|/\delta B_d$, at 10~MHz using Eqs.~(\ref{eq:bd_out2}), (\ref{eq:delBd}), and (\ref{eq:delBJ_Nb}) with $\delta B_{\text{OQS}}=10$~aT/$\sqrt{\text{Hz}}$ and $Q_{\text{eff}}=2\times10^{6}$. Figure~\ref{fig:opt_Nb} shows the calculated, normalized 2D distribution of the SNR values of the improved axion detector as a function of $N_{\text{in}}$ and $N_{\text{out}}$. Among various optimal values in the red region, we selected $N_{\text{in}}=2$ and $N_{\text{out}}=11$ as optimal values, which do not significantly increase the coils' width because of using 3-mil Nb wire.    

Based on these optimal parameters, Fig.~\ref{fig:DelBd} shows the estimated total magnetic noise of the axion detector with the quantum-limited OQS as a function of the frequency (red solid curve), indicating that the axion detector loses sensitivity at the frequency range below $\sim$1~MHz due to the dominant thermal noise of the superconducting LC circuit (blue dotted curve). On the other hand, above $\sim$1~MHz, it is limited by the OQS fundamental quantum field noise (orange dotted curve), thus enhancing the OQS field noise is the key to improve the axion detector sensitivity in this frequency range. The projected sensitivity of this optimized axion detector with $t_{\text{int}}=15$~s integration time at each observation frequency is shown in Fig.~\ref{fig:sen} (red dotted curve). We anticipate that with sensitivity up to 7 orders of magnitude beyond the current best limit, the improved axion detector with the quantum-limited OQS can potentially access the compelling targets below the KSVZ QCD axion band and probe the QCD axion parameter space in a mass range near 10$^{-8}$~eV, corresponding to the frequency range of $\sim$1~MHz.

\subsection{Mitigation of Backaction OQS Noise}
For the optimized superconducting output coil with $r_{\text{out}}=~5.0$~cm,  $d=8.0$~cm, and $N_{\text{out}}=11$, the magnetic flux of the backaction OQS noise through the output coil in Eq.~(\ref{eq:backaction}) is $\delta \Phi_{SN}=~1.9\times 10^{-21}~\text{T}\cdot\text{m}^2/\sqrt{\text{Hz}}$. For 1~MHz frequency, we compare the backaction noise $\delta V_{SN}=\delta \Phi_{SN} \omega=1.2\times 10^{-14}$~V/$\sqrt{\text{Hz}}$ with the thermal Johnson noise inside the superconducting LC circuit $\delta V_J=\sqrt{4 k_B T R_{Nb}}=3.3\times 10^{-16}$~V/$\sqrt{\text{Hz}}$, assuming $R_{Nb}=1~\text{n}\Omega$ at 2~K. This indicates that the backaction OQS noise can become significant. Hence, proper detuning of the OQS magnetic resonance can be helpful to make the two noises comparable. The OQS field noise and the backaction noise follow the Lorentzian shape, and when the detuning exceeds the resonance width they decrease as $[2\pi (\nu-\nu_0)T_2]^{-1}$. For example, a detuning from the resonance of the LC circuit by 4.6~kHz can reduce the backaction noise by 100 times, resulting in $\delta V_{SN}\approx\delta V_J$, while the OQS field noise will increase to 1~ fT/$\sqrt{\text{Hz}}$. Reaching 10 aT field noise will give some extra sensitivity for detuning to decrease the backaction and increase the axion detector bandwidth, accelerating  scanning~\cite{PhysRevD.106.112003}.

Spin squeezing can be a promising mitigation method to considerably suppress the backaction noise without sacrificing the OQS field noise. Significant reduction of the OQS spin noise by spin squeezing has been demonstrated, e.g., 70$\%$ noise reduction in Ref.~\cite{Kuzmich}, a factor of 6.4 in Ref.~ \cite{squeezingchapter}, and even a factor of 100 in Ref.~\cite{Nature100}. In fact, the OQS field noise and the backaction noise are correlated, hence conducting OQS measurements during the minimum of the oscillating spin noise can reduce both the backaction noise and the spin projection noise in Eq.~(\ref{eq:QNL}). In continuous non-demolition measurement~\cite{ShahSpinNoise}, the data can be processed to weigh signals when the spins are directed toward the output coil (i.e., parallel to its symmetry axis). This measurement will have the minimal backaction noise if at the same time the spin state is read out with the probe laser beam. During this measurement, the orthogonal spin noise component reaches the maximum; however, it is perpendicular to the output coil's symmetry axis and thus is not contributed to the backaction noise. As a result, it could be possible to suppress the backaction noise without sacrificing the OQS field noise. In contrast to the previous papers where the goal of using spin squeezing was to demonstrate the reduction in OQS field noise, here the spin squeezing will be essential for the backaction noise reduction.  
In addition, it is important to note that for a long continuous measurement the advantage of spin squeezing is removed, but for short measurements $< T_2$, the advantage can be on the order of the spin squeezing. Because of the coherence of the axion signal over 0.25~s, if we periodically implement the protocol of spin squeezing with the period matching that of the axion signal, then the averaging of multiple measurements can lead to $1/\sqrt{N}$ noise reduction, with $N$ being proportional to the measurement time, with similar improvement of the SNR to long measurements of coherent signals.

\section{Conclusion}\label{sec:conc}
We have built a prototype axion detector operating at room temperature, comprised of the optimized LC circuit and the commercial OQS with 10~fT/$\sqrt{\text{Hz}}$ field noise. The LC circuit contained the first-order one-turn planar gradiometer input coil located inside the large bore of the 2-T superconducting magnet and the two-loop four-turn circular output coil coupled to the commercial OQS sensor head. We tuned the prototype at about 300~kHz and obtained background data. We investigated the sensitivity of the prototype based on the background data and showed that the prototype experiment can probe the axion dark matter on the significant mass range between 10$^{-11}$~eV and 10$^{-7}$~eV. The sensitivity, limited by the OQS field noise of 10~fT/$\sqrt{\text{Hz}}$, is up to 1 order of magnitude better than the current best limit. We also investigated the potential sensitivity of an axion detector  based on the superconducting LC circuit, the OQS reaching the fundamental quantum noise limit of 10~aT/$\sqrt{\text{Hz}}$, and the existing 2-T magnet at LANL. The improved axion detector can potentially enhance the experimental sensitivity up to 7 orders of magnitude beyond the current best limit, allowing us to probe the QCD axion parameter space in a mass range near 10$^{-8}$~eV. The improved experiment will be limited by the quantum noise limit of the OQS. In addition, we characterized possible background noises in the experiment including the backaction OQS noise and the thermal noise of the copper shield, showing that these noises can be reduced below the thermal noise of the superconducting LC circuit.\\

\begin{acknowledgments}
The authors gratefully acknowledge the support by the Los Alamos National Laboratory LDRD office through Grants No. 20190113ER, 20210254ER, and 20230633ER. The authors are grateful for helpful discussion and assistance with operating the magnet and constructing the prototype axion detector with Dr. Leonardo Civale, Jaren Cordova, Leonard Gonzales, Larry Burney, and Shaun Newman. The authors are also grateful to Dr. Daniele Alves for useful comments.\\
\end{acknowledgments}

\bibliography{Reference}
\end{document}